\documentstyle[aps,preprint,graphicx]{revtex}

\tighten

\begin{document}

\title{Dynamics of open quantum systems}

\author{J. Oko{\l}owicz{$^{1,2}$}, M. P{\l}oszajczak{$^1$} 
and I. Rotter{$^3$}}

\address{$^1$ Grand Acc\'{e}l\'{e}rateur National d'Ions Lourds (GANIL),
CEA/DSM -- CNRS/IN2P3, BP 5027, F-14076 Caen Cedex 05, France \\
$^2$
Institute of Nuclear Physics, Radzikowskiego 152,
PL - 31342 Krak\'ow, Poland \\
$^3$
Max Planck Institute for the Physics of Complex Systems,
N\"othnitzer Str. 38,
D-01187 Dresden, Germany}

\maketitle             

\begin{abstract}

The coupling between the states of a system and the continuum
into which it is embedded, induces correlations  
that are especially large in the short time scale. 
These correlations cannot be calculated by using  a statistical
or perturbational approach. They are, however, involved in an
approach describing  structure and reaction aspects  in a
unified manner. Such a  model is the SMEC (shell model embedded in the
continuum). Some characteristic results obtained from SMEC as well as 
some aspects of the correlations induced by the coupling to the continuum 
are discussed.

\end{abstract}

\section{Introduction}

Most states of a nucleus are embedded in the contiuum of decay
channels due to which they get a finite life time. That means: the
discrete states of a nucleus shade off into resonance states with complex
energies ${\cal E}_k = E_k - \frac{i}{2} \, \Gamma_k$. The  $E_k$ give the 
positions in energy of the resonance states while the 
widths  $\Gamma_k$ are characteristic of their life times. The $E_k$ may be
different from the energies of the discrete states, and the widths 
$\Gamma_k$ may be large corresponding to a short life time. Nevertheless,
there is a well defined relation between the 
discrete states characterizing  the closed system, and the resonance states
appearing in the open system. 
The main difference in the theoretical description 
of quantum systems without and with coupling to an environment is that the
function space of the 
system is supposed to be complete in the first case
while this is not so in the second case. 
Accordingly, the Hamilton operator is Hermitian in the first case, and
the eigenvalues are discrete.
The resonance states, however, characterize a subsystem
described by a non-Hermitian Hamilton operator with complex eigenvalues. The
function space containing everything consists, in the second case, of system 
plus environment.

The mathematical formulation of this problem goes back to Feshbach 
\cite{feshbach} who
introduced the two subspaces $Q$ and $P$, with $Q+P=1$, containing the discrete
and scattering states, respectively. Feshbach was able to formulate a
unified description of nuclear reactions with both  direct
processes in the short-time scale and  compound nucleus processes
in the long-time scale. Due to the high 
excitation energy and high level
density in compound nuclei, he introduced statistical approximations 
in order to describe the
discrete states of the $Q$ subspace. A unified description of nuclear
structure and nuclear reaction aspects is much more complicated and became
possible only at the end of the last century (see \cite{rep} for a recent
review). In this formulation, the states of both subspaces are described with
the same accuracy. All the coupling matrix elements between different discrete
states, different scattering states as well as between discrete and scattering 
states are calculated in order to get results that can be compared with
experimental data.   This method has been applied to the description of light
nuclei by using the shell model approach for  the
discrete many-particle states of the $Q$ subspace \cite{rep}.

In the unified description of structure and reaction aspects, 
the system is described by an effective  Hamiltonian ${\cal H}$ that 
consists of two terms: the Hamiltonian matrix $H$ of the closed system 
with discrete eigenstates, 
and the coupling matrix  between system and environment. 
The last term is responsible for 
the finite lifetime of the resonance states. The
eigenvalues of ${\cal H}$ are complex and
give the poles of the $S$ matrix.

The dynamics of quantum systems  is determined by the $S$ matrix, 
more exactly by 
its poles and the postulation of unitarity. The  unitarity 
is involved in the continuum shell model calculations \cite{rep},
but is in conflict with the statistical assumptions when calculations in the
overlapping regime are performed \cite{guhr}.

Characteristic of the motion of the poles of the $S$ matrix as a function of a
certain parameter  are the following generic results
obtained for very different systems 
\cite{marost,ro01,pepirose,napirose}:
in the overlapping regime, the trajectories 
of the $S$ matrix poles avoid crossing with the only exception 
of exact crossing
when the $S$ matrix has a double  pole. At the avoided crossing, 
either level repulsion or level attraction occurs. The first case is caused by
a predominantly real interaction between the crossing states and is 
accompanied by the tendency
to form a uniform  time scale of the system. Level attraction occurs,
however, when the interaction is dominated by its imaginary part 
arising from the coupling via the continuum. It is
accompanied by the formation of different time scales in the system: while
some of the states decouple more or less completely from the continuum 
and become
long-lived (trapped), a few of the states become short-lived and wrap
the long-lived ones in the cross section.
The dynamics of  quantum systems at high level density is determined 
by the interplay of
these two opposite tendencies. For a more detailed discussion see \cite{rep}.

At large overall coupling strength, quick direct reaction processes may appear 
from slow resonance processes by means of the resonance trapping phenomenon.
In recent calculations on
microwave cavities with fixed  large overall coupling strength, 
the details of resonance trapping are shown to depend 
on the position of the attached leads \cite{napirose}. 
In microwave cavities of Bunimovich type, that are chaotic when closed,
coherent whispering gallery and bouncing ball modes may be strongly enhanced
by trapping other (incoherent) modes.
Since the coherent modes determine the value of the conductivity,
resonance trapping may cause  observable effects that are not small.
It is interesting to remark that the trapped long-lived states can be
described by random matrix theory, as a shot-noise analysis of the numerical
results has shown. The enhanced conductivity, however, is related
to the  short-lived whispering gallery modes that are regular and dominant
when the leads are attached to the cavity in a suitable manner \cite{shot}.

Meanwhile, the phenomenon of  resonance trapping has been
proven experimentally on a microwave cavity as a function of the degree of
opening of the cavity to an attached lead  \cite{stm1}.  
In this experiment, the parameter varied is the overall
coupling strength between discrete and scattering states. 
Resonance trapping may appear, however,  as  function of any parameter
\cite{rep}. 

In the following, we will discuss the interplay of the different time scales 
in nuclei. Most interesting is the mechanism of formation of short-lived
states in open quantum systems. In Sect. 2,
the effective Hamiltonian and the $S$
matrix are written down for a quantum system embedded in a continuum while
in Sect. 3, the basic relations for the spectroscopic information are
discussed. Characteristic features of the different approaches to the system 
and to the  environment are sketched  in Sect. 4.
In Sect. 5, some results obtained from calculations with unified description
of structure and reaction aspects are shown. The relation between lifetimes
and decay widths of resonance states in the overlapping region is discussed in
Sect. 6 while in Sect. 7,   properties of the system in the
short time scale are illustrated. In any case, the correlations induced by the
coupling to the continuum are large. The last section contains some concluding
remarks.

\section{Effective Hamiltonian and 
$S$ matrix for a quantum system embedded in a continuum}

In the unified description of structure and reaction aspects of quantum 
systems, the Schr\"odinger equation 
\begin{eqnarray} 
(H-E)\Psi = 0
\label{csm1}
\end{eqnarray}
is solved in
a function space containing everything, i.e. discrete as well as continuous
states. The Hamilton operator $H$ is Hermitian, the wave functions $\Psi$ 
depend on energy as well as on the decay channels and all the resonance
states of the system. Knowing the wave functions 
$\Psi$, an expression for the $S$ matrix can be derived
that holds true also in the overlapping regime,
see the recent review \cite{rep}. In the continuum shell model, it reads
\begin{eqnarray} 
 S_{cc'} = e^{i(\delta_c-\delta_{c'})}\; \Big[
 \delta_{cc'} -  S_{cc'}^{(1)} -  S_{cc'}^{(2)} \Big] ~ \ ,
\label{csm2}
\end{eqnarray}
where $ S_{cc'}^{(1)}$
is the smooth direct reaction part related to the short-time scale,
and 
\begin{eqnarray}
 S_{cc'}^{(2)} =  i \; \sum_{k=1}^N
  \frac{\tilde\gamma_{k}^{c}\; \tilde\gamma_{k}^{c'}}
  {E - {\tilde E}_k + \frac{i}{2} {\tilde \Gamma}_k} 
\label{csm3}
\end{eqnarray}
is the resonance reaction part related to the long-time scale.
Here, the   $\tilde{\cal E}_k = \tilde E_k - \frac{i}{2}\; 
\tilde\Gamma_k$ are the complex eigenvalues 
of the non-Hermitian Hamilton operator 
\begin{eqnarray}
{\cal H}_{QQ} = H_{QQ} + H_{QP} \, G_P^{(+)} \, H_{PQ}
\label{csm4}
\end{eqnarray}
appearing effectively in the system ($Q$ subspace)
after embedding it into the continuum ($P$ subspace). They are 
energy dependent functions and determine  the positions $E_k = 
\tilde E_k ${\footnotesize$\, (E\!\!=\!\! E_k)$}
and widths $\Gamma_k = \tilde \Gamma_k ${\footnotesize$\, (E\!\!=\!\! E_k)$} 
of the resonance states $k$ \cite{rep}. 
The $G_P^{(+)}$ in (\ref{csm4}) are the Green functions in the $P$ subspace. 
The $\tilde \gamma_{k}^{c}$  are the coupling matrix elements
between the resonance states and the 
scattering states. They are also energy dependent functions.
The wave functions $\tilde\Omega_k$ of the resonance states 
are related to the eigenfunctions $\tilde \Phi_k$ of ${\cal H}_{QQ}$ by a
Lippmann-Schwinger like relation \cite{rep},
\begin{eqnarray}
\tilde \Omega_k = (1+  G_P^{(+)} H_{PQ}) \, \tilde \Phi_k
\; .
\label{csm4a}
\end{eqnarray}
The eigenfunctions of ${\cal H}_{QQ}$
are bi-orthogonal,
\begin{eqnarray} 
\langle \tilde \Phi_l^* | \tilde \Phi_k \rangle = \delta_{kl} 
\label{csm5}
\end{eqnarray}
so that
\begin{eqnarray}
\langle \tilde \Phi_k|\tilde\Phi_k \rangle =
{\rm Re} ( \langle \tilde\Phi_k|\tilde\Phi_k \rangle) 
 \; \;  & ; & \; \; A_k \equiv   \langle \tilde \Phi_k|\tilde\Phi_k \rangle
\ge 1             
 \label{csm6}
\\
 \langle \tilde\Phi_k|\tilde\Phi_{l\ne k} \rangle  =  
i \; {\rm Im} (\langle \tilde\Phi_k|\tilde\Phi_{l\ne k} \rangle )  
 =  -\langle \tilde\Phi_{l\ne k}|\tilde\Phi_k \rangle 
\; \; & ; & \; \; B_k^{l\ne k} \equiv  
| \langle\tilde \Phi_k|\tilde\Phi_{l\ne k} \rangle| \ge 0 ~ \ . 
  \label{csm7}
\end{eqnarray}
As a consequence of (\ref{csm6}), it holds \cite{rep}
\begin{eqnarray}
\tilde \Gamma_k = \frac{\sum_c |\tilde \gamma_k^c|^2}{ A_k}
\; \le \;  \sum_c |\tilde \gamma_k^c|^2 ~ \ .
\label{csm8}
\end{eqnarray}

The main difference to the standard theory is that
the $\tilde\Gamma_k, \; \tilde \gamma_k^c$ and $\tilde E_k$
are not numbers but  energy dependent functions
\cite{rep}.  The energy dependence of  
${\rm Im}\{\tilde {\cal E}_k\} =  - \frac{1}{2} \, \tilde \Gamma_k$
is large near to the  threshold for opening the first decay channel. 
This causes not only deviations from the Breit Wigner line shape
of isolated resonances lying near to the threshold, but also an 
interference with the above-threshold ''tail'' of bound states,  
see Sect. 5.2 for an example. 
Also an inelastic threshold may have an influence on the line shape
of a resonance when the resonance  lies near to the threshold and  
is coupled strongly to the channel which opens \cite{ro91}.
Also in this case, $\tilde \Gamma_k$ depends strongly on energy.  
In the cross section, a cusp may appear 
in the cross section instead of a resonance  of Breit Wigner  shape. Both
types of threshold effects in the line shape of resonances  
can explain experimental data known in nuclear physics \cite{rep}.
They can not be simulated by a parameter in the $S$ matrix.

In the numerical calculations in the framework of the continuum shell model,
the coupling matrix elements $\tilde \gamma_k^c$ between resonance states and
continuum are obtained by representing the  
eigenfunctions $\tilde \Phi_k$ of the effective  non-Hermitian Hamilton 
operator $\tilde {\cal H}_{QQ}$ in the set of eigenfunctions 
$\{\Phi_k\}$ of the Hermitian Hamilton operator  $H_{QQ}$ , 
\begin{eqnarray}
\tilde \Phi_k = \sum_l b_{kl} \, \Phi_l \; .
\label{csm9}
\end{eqnarray}
The  $\Phi_k$ are real, while the  $\tilde \Phi_k$ are complex
and energy dependent. The
coefficients $b_{kl}$ and   the $(\tilde \gamma_k^c)^2$ 
are complex and energy dependent, too. The $(\tilde\gamma_{k}^{c})^2$ 
characterizing the coupling of the resonance state $k$ to the continuum, 
are related to the width of this state. In the overlapping regime,
their sum over all channels is, however, not equal to the width even in the 
one-channel case, eq. (\ref{csm8}). 
Both functions,  $(\tilde\gamma_{k}^{c})^2$ and $\tilde \Gamma_k$,
may show  a different energy dependence. An example is shown in Sect. 5.3.

\section{Spectroscopy of resonance states}

\subsection{Isolated  resonance states}

The energies and widths 
of the resonance states follow from the solutions of the fixed-point
equations :
\begin{eqnarray}
E_k = \tilde E_k
{\mbox{\footnotesize $(E\!\!=\!\!E_k)$}} 
\label{eq:fixp1}
\end{eqnarray}
and
\begin{eqnarray}
\Gamma_k  =  \tilde \Gamma_k
 {\mbox{\footnotesize $(E\!\!=\!\!E_k)$}} ~ \ ,
\label{eq:fixp2}
\end{eqnarray}
on condition that the two subspaces are defined adequately \cite{rep}.
The values $E_k$ and $ \Gamma_k$ 
correspond to the standard spectroscopic observables. The functions 
$\tilde E_k(E)$
and $\tilde \Gamma_k(E)$ follow from the eigenvalues $\tilde {\cal E}_k$
of  ${\cal H}_{QQ}$. The wave functions of the resonance 
states are defined by the functions 
$\tilde \Omega_k$, Eq. (\ref{csm4a}), at the energy $E=E_k$.
The partial widths are related to the coupling matrix elements 
$(\tilde \gamma_{k}^{c})^2$ that are calculated  independently by means 
of the   eigenfunctions $\tilde \Phi_k$ of ${\cal H}_{QQ}$. 
For isolated resonances, $A_k=1$ according to (\ref{csm6})  and
$(\tilde\gamma_{k}^{c})^2 = |\gamma_k^c|^2$. In this case
 the standard relation 
$\Gamma_k = \sum_c |\gamma_{k}^{c}|^2 $ follows from (\ref{csm8}).

It should be underlined 
that different  $\tilde \Phi_k {\mbox{\footnotesize $(E\!\!=\!\!E_k)$}}
$ are neither strictly orthogonal nor bi-orthogonal since 
the bi-orthogonality relation (\ref{csm5})
holds only when the  energies of both states $k$ and $l$ are equal. 
The spectroscopic studies on resonance states are performed therefore with the
wave functions being only approximately bi-orthogonal. The deviations 
from the bi-orthogonality
relation (\ref{csm5}) are small, however, since the $\tilde \Phi_k$ 
depend only weakly on the energy. 

This drawback of the spectroscopic studies of resonance states has to be
contrasted with  the  advantage it has for the study of observable values: 
the $S$ matrix and therefore the cross section is calculated with the
resonance wave functions being strictly bi-orthogonal at every energy $E$ of
the system. Furthermore, the full energy dependence of $\tilde E_k, \tilde
\Gamma_k$ and, above all, of the coupling matrix elements $\tilde \gamma_k^c$
is taken into account in the $S$ matrix and therefore in all calculations for
observable values.

As a result of the formalism sketched in Sect. 2
for describing the nucleus as an open quantum system,
the influence of the continuum of scattering states on the spectroscopic
values consist mainly in the following:  
there is (i) an additional shift in energy of the states
and (ii)  an additional mixing of the states
through the continuum of decay channels.

For  isolated resonances, the additional shift
is usually taken into account by simulating  
Re$({\cal H}_{QQ}) = H_{QQ} + {\rm Re}\, (W)$  (see Eq. (\ref{csm4}))
by $H_0 + V'$,  where $V'$ contains  the two-body effective 
residual forces and $W\equiv H_{PQ} G_P^{(+)} H_{PQ}$.
Furthermore, the widths of isolated states are not  calculated from
${\rm Im}\, (W)$, but from the sum of the partial widths. The
amplitudes of the partial widths are the 
coupling matrix elements between the discrete states of the
$Q$ subspace  and the scattering wave functions of the $P$ subspace. 
The additional mixing of the states  via the continuum 
is neglected in the standard calculations.

It should be mentioned, however, that  Re$(W)$ can not 
completely be simulated 
by an additional contribution to the residual 
two-body interaction since it contains many-body effects,
as follows from the analytical structure of $W$.
Re$(W)$ is an integral over energy 
and depends explicitly on the energies $\epsilon_c$ at which  the
channels $c$ open. As a matter of fact, the thresholds for neutron and 
proton channels in nuclei open at different energies. Therefore,  Re$(W)$ 
causes some charge dependence of the effective nuclear forces 
in spite of the charge symmetry of the 
Hamiltonian $H_{QQ}$. It arises as a many-body effect
depending on shell closures, and is directly 
related to the different binding energies of neutrons and protons in 
nuclei \cite{rep,ro91}.

Since only a few data on isolated resonances are  
sensitive to the many-body effects involved in Re$(W)$,
the standard calculations performed by using a Hermitian operator
are mostly justified.  However, the standard calculations can not 
be  justified for 
closely-lying levels which are coupled via the continuum of decay channels,
as well as for well isolated levels in the neighbourhood 
of thresholds where new decay channels open.

\subsection{Correlations induced by the coupling via the continuum}

The coupling of the resonance states via the continuum 
induces correlations between the states that are described by
the term $H_{QP}G_P^{(+)}H_{PQ} \equiv W$ of the effective 
Hamiltonian ${\cal H}_{QQ}$, Eq. (\ref{csm4}). 
$W$ is complex  and energy dependent \cite{rep}. 
The real part Re$(W)$ causes level repulsion
in energy and is accompanied by the tendency to form a uniform time scale 
in the system. In contrast to this behaviour, the imaginary part 
Im$(W)$ causes different time scales 
in the system and is accompanied by level attraction in energy.
That means, the formation of correlations at short-time scales is 
essentially influenced by Im$(W)$.

In the overlapping regime, many calculations have shown the phenomenon of
resonance trapping caused by Im$(W)$,
\begin{eqnarray}
\sum_{k=1}^N \tilde \Gamma_k \approx \sum_{K=1}^K \tilde \Gamma_k \quad ;
\quad \quad \sum_{k=K+1}^N \tilde \Gamma_k \approx 0 
\; .
\label{corr1}
\end{eqnarray}
It means almost complete decoupling of $N-K$ resonance states from the
continuum while $K$ of them become short-lived. Usually, $K\ll N-K$. The
long-lived resonance states in the overlapping regime appear often to be well
isolated from one another. The few short-lived resonance states determine the
evolution of the system (short time scale).

The formation of different time scales
in an open quantum system that is accompanied by level attraction, is
accompanied also by the appearance of a non-trivial 
energy dependence of the $W$ \cite{ro03}. This energy dependence 
can directly be expressed by non-linear terms appearing in the
overlapping regime. 
As a consequence, the  use of an effective Hamiltonian in describing 
scattering processes is meaningful only when, at the same time, the
energy dependence of the  $W$ is considered.

In the framework of statistical approaches, 
the  coupling matrix elements between
resonance states and continuum  are assumed to be parameters
being  energy independent.
Also in the different versions of $R$ matrix approaches,
the correlations induced by $W$ cannot be studied.
The interplay between the different time scales of open quantum systems 
at high level density can
be studied only microscopically, without any 
statistical assumptions on the level distribution or perturbation 
theory approaches.

\subsection{Overlapping resonance states}

The solutions $E_k$ and $\Gamma_k$
of the fixed point equations (\ref{eq:fixp1}) and (\ref{eq:fixp2})
are  basic  for spectroscopic studies not only of isolated but also 
of overlapping resonances since the energy dependence of the 
eigenvalues $\tilde {\cal E}_k = \tilde E_k - i/2 \; \tilde \Gamma_k$ 
of the effective Hamiltonian ${\cal H}_{QQ}$ 
is smooth everywhere.  The $E_k$ and $\Gamma_k$ are therefore well defined
and it makes  sense to use them for spectroscopic studies.  
The coupling coefficients $\tilde \gamma_k^c$  are however worse defined 
since the wave functions
$\tilde \Phi_k {\mbox{\footnotesize $(E\!\!=\!\!E_k)$}}$ 
are bi-orthogonal.  The 
bi-orthogonality relations (\ref{csm6}) and (\ref{csm7})
become important at the avoided level crossings where $A_k > 1$.
In approaching a double pole of the $S$ matrix,  
$A_k \to \infty$. The same holds for the 
modulus square of the coupling coefficients:
$|\tilde\gamma_k^c|^2 \to \infty$, in accordance with  the relation
(\ref{csm8}). 

The numerator of the resonance part of the  $S$ matrix (\ref{csm3}) is 
\begin{eqnarray}
\langle \tilde \Phi_k^* | \hat W_{cc'} | \tilde \Phi_k \rangle 
= 2 \pi  
\langle \tilde \Phi_k^* | V^\dagger | \xi_E^{c} \rangle 
\langle \xi_E^{c'} | V | \tilde \Phi_k \rangle
=  \tilde \gamma_k^c  \tilde \gamma_k^{c'}
~ \ .
\label{eq:biorth21}
\end{eqnarray}
For $c = c'$, this is $(\tilde \gamma_k^c)^2$  and not
$|\tilde\gamma_k^c|^2 $ as often assumed  \cite{mawei}. 
Expression  (\ref{eq:biorth21}) 
remains  meaningful also in approaching the double pole of the $S$
matrix \cite{rep}, and the $S$ matrix (\ref{csm2}) with (\ref{csm3}) 
is unitary also in the overlapping regime. When the energy difference
$\Delta E=|E_k - E_l|$ between two neighbouring resonance states
is smaller than their widths,
higher-order terms in the $S$ matrix that are related to the 
bi-orthogonality of
the eigenfunctions of the non-Hermitian Hamilton operator ${\cal H}_{QQ}$,
can not be neglected.
At a double pole of the $S$ matrix, 
$ (\tilde \gamma_k^c)^2  \to - (\tilde \gamma_l^c)^2 $  
corresponding to $\tilde \Phi_k \to \pm \, i \, 
\tilde \Phi_l $ \cite{rep}.
Here, the two resonance terms cancel, and the system decouples from the
continuum at the energy of the double pole. The same relations hold 
when the two states avoid crossing in the complex plane
by varying a certain parameter \cite{ro01}.
The point is, however, that in such a case 
the transition $\tilde \Phi_k \to \pm \, i \, \tilde \Phi_l $  influences 
the wave functions not only at the critical point but   in  a certain
region around the critical value of the parameter \cite{ro01}.  
At high level density, this fact will cause deviations from
the relation $ \tilde \Gamma_k  = \sum_c (\tilde \gamma_k^c)^2 $.
For numerical results on the relation between $ \tilde \Gamma_k$
and $(\tilde \gamma_k^c)^2 $, see Sect. 5.3.

Furthermore,
the energies and widths of overlapping resonance states 
are given by the values $E_k$ and $\Gamma_k$ (Eqs. (\ref{eq:fixp1})
and (\ref{eq:fixp2})), 
at which the $S$ matrix has poles. However, the positions of the maxima 
in the cross section do, generally,
not appear at the energies $E_k$ when the resonance states overlap
\cite{rep}. 

The relation between  $\tilde \Gamma_k = - \, 2 \;
{\rm Im}\; \{\langle \tilde \Phi_k^* | {\cal H}_{QQ} |
\tilde  \Phi_{k} \rangle\} $  and the sum of the coupling coefficients
$\sum_c (\tilde \gamma_k^c)^2 $ 
is, in general, more complicated than for isolated resonances
due to the avoidance of level crossings in the complex plane 
\cite{rep}. The $S$ matrix 
behaves smoothly in the neighbourhood of a double pole. The same is true 
for measurable values due to their relation to the $S$ matrix. 
The value $ |\tilde \gamma_k^c|^2$ loses its physical meaning in the 
overlapping regime.

\section{Different approaches}

\subsection{Statistical approach to the system}

More than 40 years ago, the {\it unified theory of nuclear reactions} has been
formulated by Feshbach \cite{feshbach}. Feshbach introduced the projection
operator technique in order to make possible the  concurrent 
numerical solution of equations with    
discrete and scattering states in spite of their very different 
mathematical properties. By means of the projection operator technique, the
whole function space is divided into the subspace of discrete states ($Q$
subspace) and the subspace of scattering states ($P$ subspace).
Then, the problem in the $P$ subspace is solved numerically by coupled-channel 
methods  while the problem in the $Q$
subspace is not solved directly. Here, statistical assumptions are introduced
by which the mean properties of the discrete states are described. Also the
coupling matrix elements between discrete and scattering states are 
determined statistically and characterized by their mean values. 

The advantage of using different approximations  in the two subspaces
consists, above all, in the possibility to solve the coupled-channel problem
with high accuracy. Since the  $P$ subspace is constructed from all 
open decay channels, it changes with energy
since new  channels  open in passing the corresponding thresholds. Furthermore,
the  inclusion of, e.g., $\alpha$ decay channels into the $P$ subspace is 
not a problem. The method is applied successfully to the description of nuclear
reactions in energy regions with high level density of the excited nucleus
which makes it possible for a statistical treatment of the discrete states of 
the $Q$ subspace. It represents the standard method in analyzing nuclear
reaction data on medium and heavy nuclei at low energy.

The {\it shell model approach to nuclear reactions} \cite{mawei} is 
formulated by Mahaux and Weidenm\"uller. 
Also in this approach, the whole function space is divided into the two
subspaces. However, the $P$ subspace contains open as well as closed decay
channels and, therefore, does not change with energy. The inclusion of more
than one particle in the continuum becomes a principal problem.
The bi-orthogonality of the eigenfunctions of the effective Hamiltonian 
is not considered what causes problems with the unitarity of the $S$ matrix in
the overlapping regime due to $\Gamma_k < \sum_c |\gamma_k^c|^2$
\cite{mawei}. Eventually, the states of the 
$Q$ subspace  are treated by means of statistical methods in
the same manner as in the Feshbach formulation \cite{feshbach}. 
The restrictions in the applicability of both treatments
are therefore the same: as long as the (long-lived) resonance states are
isolated from each other and their individual properties 
can be neglected to a good approximation, the method  gives 
reliable results. 

The formation of different time scales in a realistic system cannot be 
studied by using a statistical description of the states, since
the interplay between the real and imaginary parts of the interaction 
in the effective Hamiltonian ${\cal H}_{QQ}$ is not
taken into account.

\subsection{$R$ matrix approach}

In contrast to the Feshbach unified theory of nuclear reactions, different
approaches for the description of decaying states are worked out 
by starting from well established nuclear structure models.
These approaches
are based on the $R$ matrix theory of nuclear reactions
that is justified at low level density \cite{lanethom}.
Here, the resonance levels are assumed to be isolated, {\it i.e.} 
the influence of resonance overlapping on the nuclear structure
is not considered.

The advantage of these studies consists, above all, in the integration of 
proven nuclear structure models into the calculations. That means, 
the wave functions of the $Q$ subspace are realistic.  
The coupling to the supplementary $P$ subspace
(continuum of decay channels) is described in a straightforward manner. The  
feedback from the continuum of decay channels on the nuclear structure
is however hidden, if at all taken into account, in the 
results of the numerical studies. 
When the resonances overlap, some averaging over many levels is  
performed in the $R$ matrix
theory of nuclear reactions \cite{lanethom}. 

The formation of different time scales in the system cannot be studied
since it arises from the  feedback from the continuum to
the states of the system that is not taken into account
in the $R$ matrix approach.

\subsection{Shell model approach to the system}

In reactions on light nuclei and in studying nuclei near to the drip line, 
the level density is low and the individual
properties of the nuclear states can not be neglected. In these nuclei, 
the restriction
to a description of the mean properties of the states is not justified. The 
problem in the $Q$
subspace has to be solved with a higher accuracy.

The spectroscopic properties of
light nuclei are described successfully in the framework of the shell model.
It is therefore reasonable to identify the $Q$ subspace with the function
space of the shell model used in performing numerical calculations for these
nuclei.  Two different approaches have been developed: (i) the 
CSM-FDP approach (continuum shell model  with finite
depth potential), that  generates the single particle basis 
states  in a Woods-Saxon potential \cite{ro91,baroho}, and has
been used mainly for a description of giant resonances  
in light nuclei, and (ii) the SMEC (shell model embedded in the continuum) 
which uses the shell model effective interaction in
the $Q$ subspace and provides, in particular, a realistic description 
of resonance phenomena near particle decay thresholds \cite{bnop}. 
Common to both approaches is that Eq. (\ref{csm1})  
is solved numerically by using similar approximations in the two subspaces. 
The bi-orthogonality of the eigenfunctions of the effective
Hamiltonian (Eqs. (\ref{csm5})
to (\ref{csm7})) is taken into account in both approaches. As a
consequence,   the unitarity of the $S$ matrix
is ensured also in the overlapping regime.
These calculations provide  a {\it unified description  of nuclear
structure and nuclear reaction} aspects. 

In the SMEC, the nuclear shell model is involved what makes it possible 
for a realistic
description of the nuclear structure, as in the models based on the 
$R$ matrix approach. However, 
in contrast to these models, the
feedback  from the continuum of decay channels on the nuclear structure
is explicitly taken into account.
Therefore, the formation of different time scales in the system 
can be studied by means of SMEC.

\section{Some results obtained for $^{24}$Mg in SMEC}

\subsection{The $^{24}$Mg nucleus}

Let us consider  $^{24}$Mg  with the inner core  $^{16}$O and the
phenomenological $sd$-shell interaction among the valence 
nucleons.
Within  this  configuration space, the $^{24}{\rm Mg}$ nucleus 
has  325 states with $J^{\pi}=0^+$, $T=0$. 
These states can couple to a number of open channels which 
correspond to excited states in the neighboring ($A-1$) nucleus.  
For details see \cite{rep,drokplro}.

\begin{figure}[h]
\begin{center}
\includegraphics[width=16cm,angle=0]{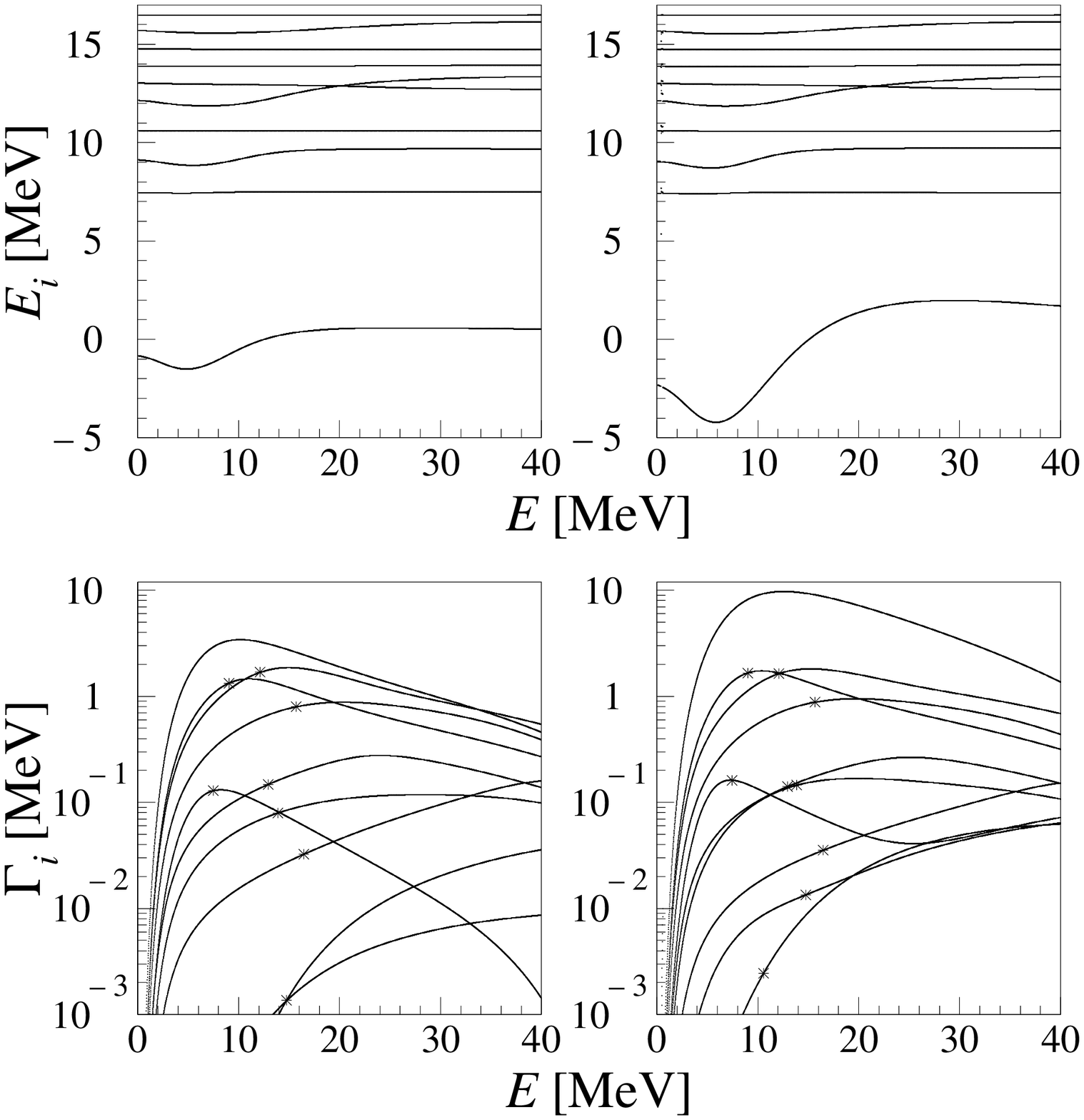}
\end{center}
\vspace*{1cm}
\caption{Energy dependence of the 
positions $\tilde E_i$ (upper row) and  widths $\tilde \Gamma_i$ (lower row)
of the ten 
lowest $0^+$ states of $^{24}$Mg as a function of the energy $E$ of the
particle in the continuum.
In the first column,
the  calculations include coupling to only one channel in $^{23}$Mg. In
the  other column,  two  channels are taken into account. 
The stars at the trajectories mark the fixed-point solutions.
}
\label{paren1}
\end{figure}

For illustration, we show in 
Fig.~\ref{paren1} the dependence of energies
$\tilde E_k$ and  widths $\tilde \Gamma_k$ of the ten 
lowest $0^+$ states of $^{24}$Mg on the energy $E$ of the
particle in the continuum, as well as the eigenvalue picture with the energy
$E$  parametrically varied. The number of channels is one (the l.h.s
plot) and two (the r.h.s. plot). We can see the non-random features 
occurring at 
this edge of the spectrum. The coupling between the channels 
reduces the differences between the widths of the different states.
It has almost no influence onto their positions.

The positions $\tilde E_k$ of the resonance states 
are almost independent of a variation of the energy of the system
(Fig.~\ref{paren1}).
The widths $\tilde \Gamma_k$ however depend on energy: they rise at low
energies above the particle decay threshold 
and decrease again at energies beyond the positions $E_k$ of the states. 
Most of the resonances have therefore a tail at the high energy side. 
This feature is well pronounced especially for the 
lowest-lying state which is  bound.
Due to its large width, it can contribute to the 
cross section in the threshold region.

\subsection{Near-threshold behaviour of the cross section}

\begin{figure}
\begin{center}
\includegraphics[width=12cm,angle=0]{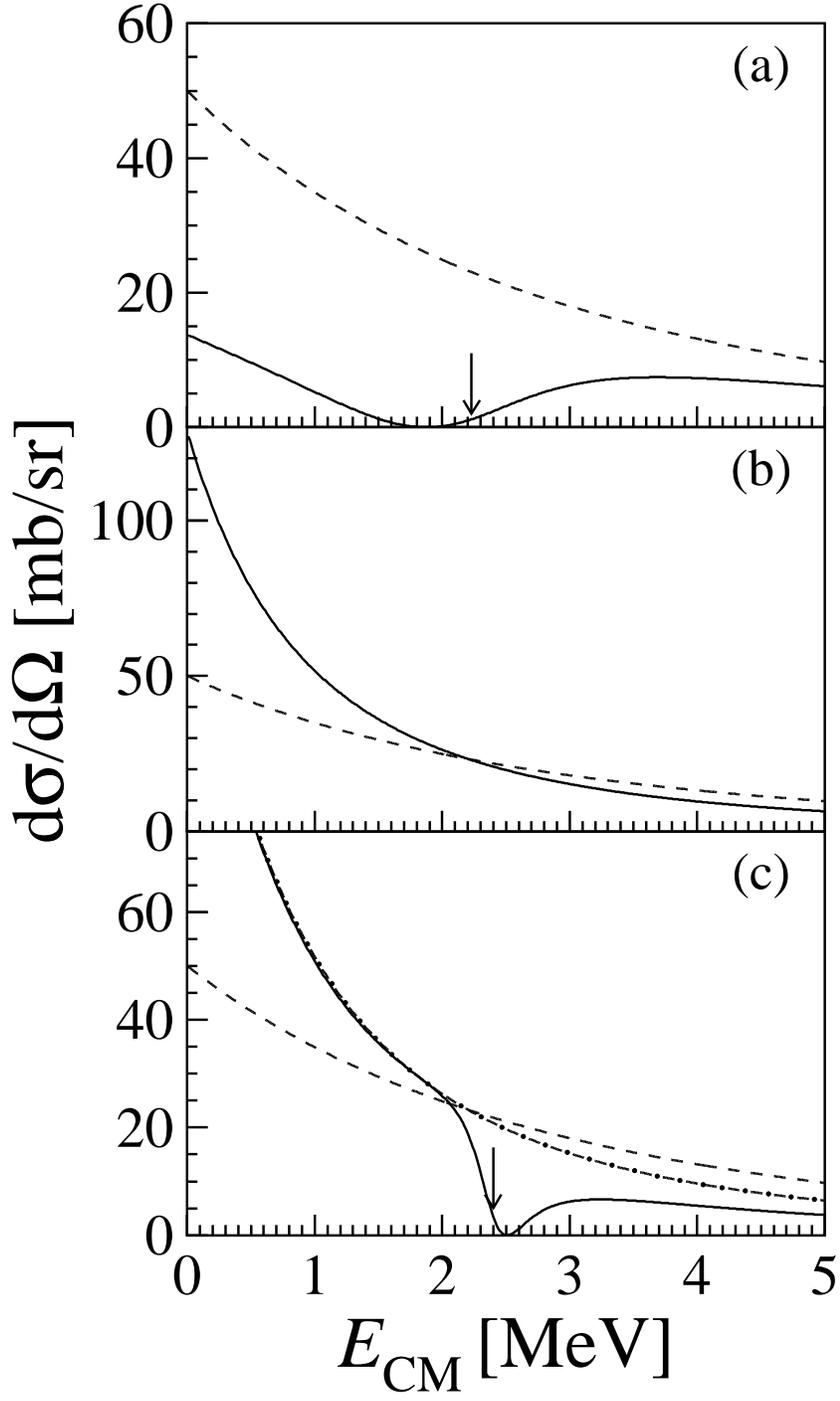}
\end{center}
\caption{
Cross section for the reaction ~n + $^{23}$Mg$ ~\to ~^{24}$Mg 
calculated for one open neutron channel and 
(a) the resonance state $0_2^+$ of $^{24}$Mg, 
(b) the ground state $0_1^+$ of $^{24}$Mg, 
(c) the bound $0_1^+$ and the
resonance state $0_2^+$ of $^{24}$Mg. The dashed lines show
the direct reaction part of the cross section. The arrows denote the position
of the resonances.
} 
\label{fig:elthr}
\noindent
\end{figure}

In Fig. \ref{fig:elthr}, we summarize the generic features of cross 
sections near thresholds.  As an example we show the cross section for the
reaction 
${\rm n}+{}^{23}{\rm Mg}\,(1/2^+) \longrightarrow {}^{24}{\rm Mg}(0^+)$ 
(one open channel, the $s$-wave scattering). 
The upper part shows the cross section with 
only one excited (resonance) state $0_2^+$ of $^{24}{\rm Mg}$.
The minimum in the cross section is an effect of destructive 
interference of the resonance ($E^*=2.23$~MeV, $\Gamma=1.76$~MeV) with
the background of the potential scattering
(the direct part of the reaction cross section),  denoted by the dashed line. 
In the middle part, the cross section with only 
the ground state (bound state) $0_1^+$ in $^{24}{\rm Mg}$  is shown.
The cross section exhibits a strong increase for $E \to 0$.
This is caused by the bound state for which $\tilde \Gamma_k \ne 0$ 
at  $E > 0$ in spite of $\tilde \Gamma_k = 0$ at  $E = \tilde E_k < 0$.   
Finally, the cross section with 
both ground state $0_1^+$ and  resonance state $0_2^+$ is shown 
in the bottom part of the figure. 
The interference picture of these two states shows level repulsion accompanied
by a decrease of the width of the higher-lying state ($E^*=2.40$~MeV,
$\Gamma=0.47$~MeV). 
The line shape of the resonance resembles a typical interference picture for
overlapping resonances in spite of the fact that the calculation is performed
with only one resonance state while the other state is bound.

\subsection{Relation between total and partial widths}

\begin{figure}[h]
\begin{center}
\includegraphics[width=16cm,angle=0]{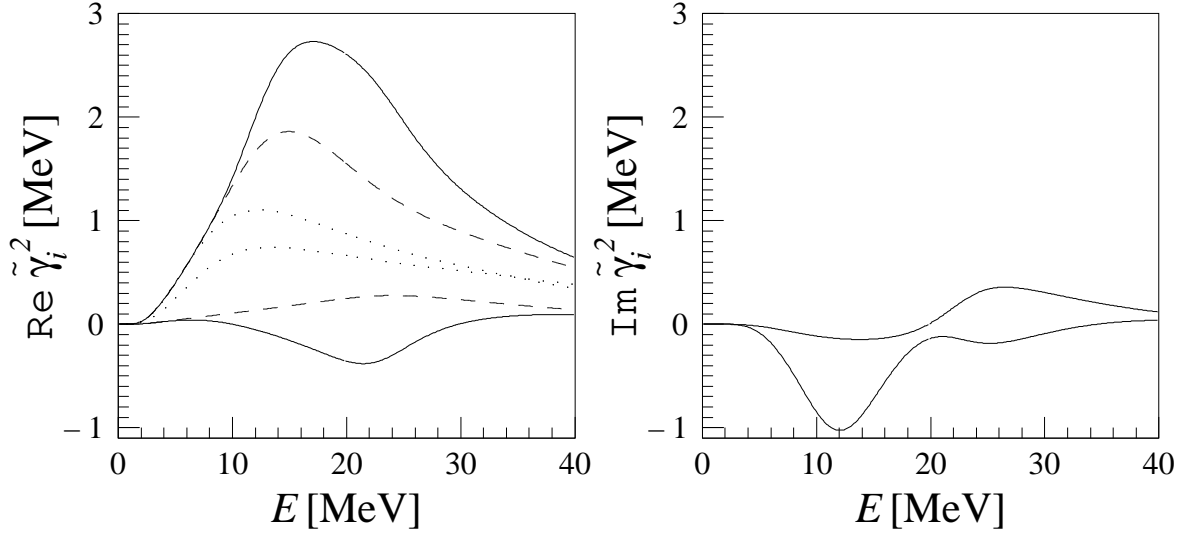}
\end{center}
\vspace*{1cm}
\caption{
Energy dependence of the coupling coefficients 
$({\tilde {\gamma}_i})^2$  (solid lines) 
for crossing resonances in $^{24}$Mg
(resonances 5 and 6 in Fig.~\protect\ref{paren1}, one-channel case). 
The real parts are shown on the left-hand side and the imaginary
parts on the right-hand side.
In addition, the figure on the left-hand side 
shows the dependence of the widths ${\tilde \Gamma}_i$ (dashed lines) and
$|{\gamma_i}|^2$ (dotted lines) on the  energy of the
particle in the continuum.
}
\label{croresen}
\end{figure}

\begin{figure}[h]
\begin{center}
\includegraphics[width=16cm,angle=0]{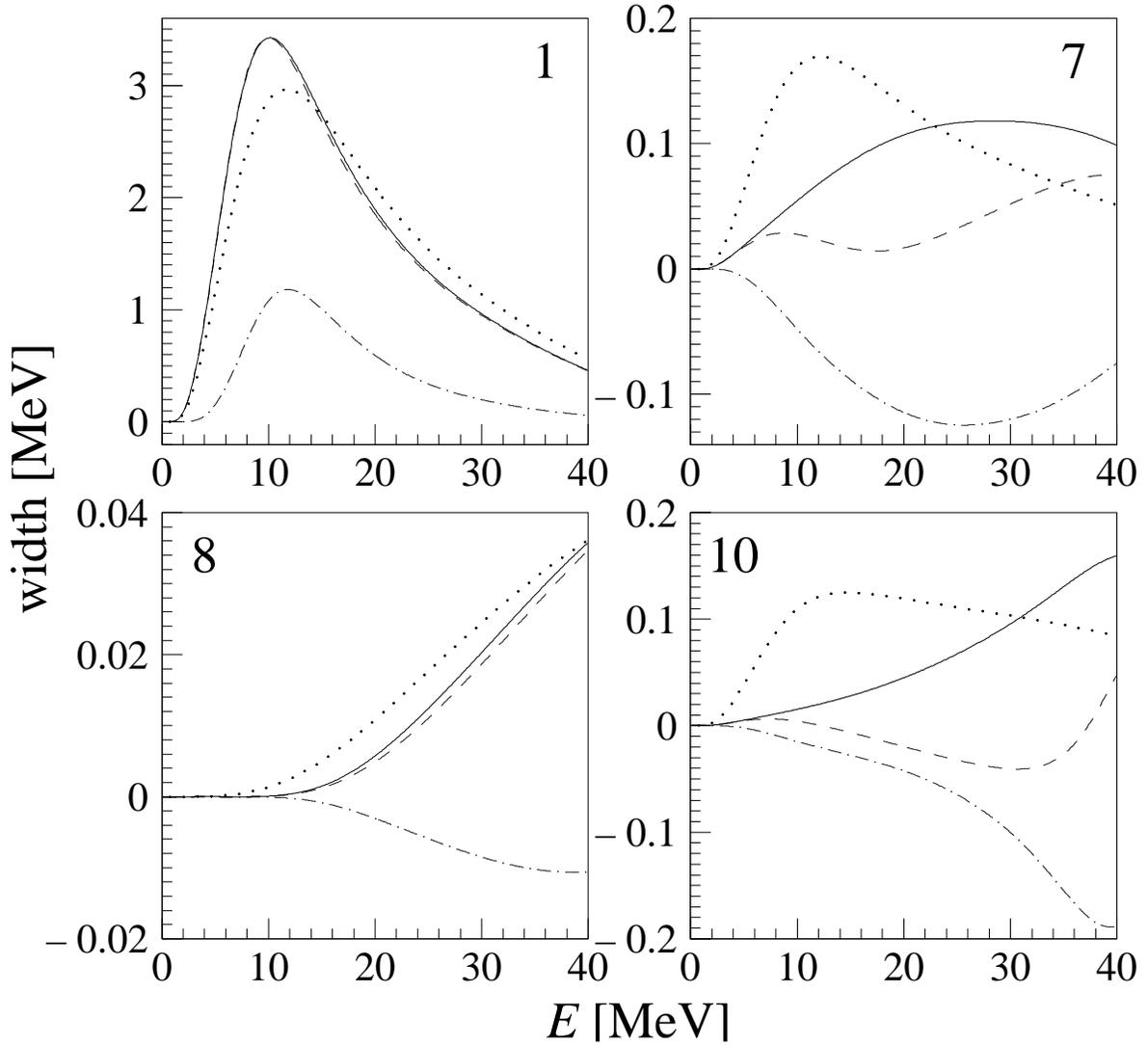}
\end{center}
\vspace*{1cm}
\caption{
Energy dependence of various `widths' 
for resonances `1', `7', `8', `10' (see  Fig.~\protect\ref{paren1})
in $^{24}$Mg. The different lines denote :  
${\tilde \Gamma}_i$ (solid line), 
Re$({\tilde {\gamma}_i})^2$ (dashed line), Im$({\tilde {\gamma}_i})^2$ 
(dashed-dotted line) and $|{\gamma_i}|^2$ (dotted line).
}
\label{croresen1}
\end{figure}

\begin{figure}[h]
\begin{center}
\includegraphics[width=12cm,angle=0]{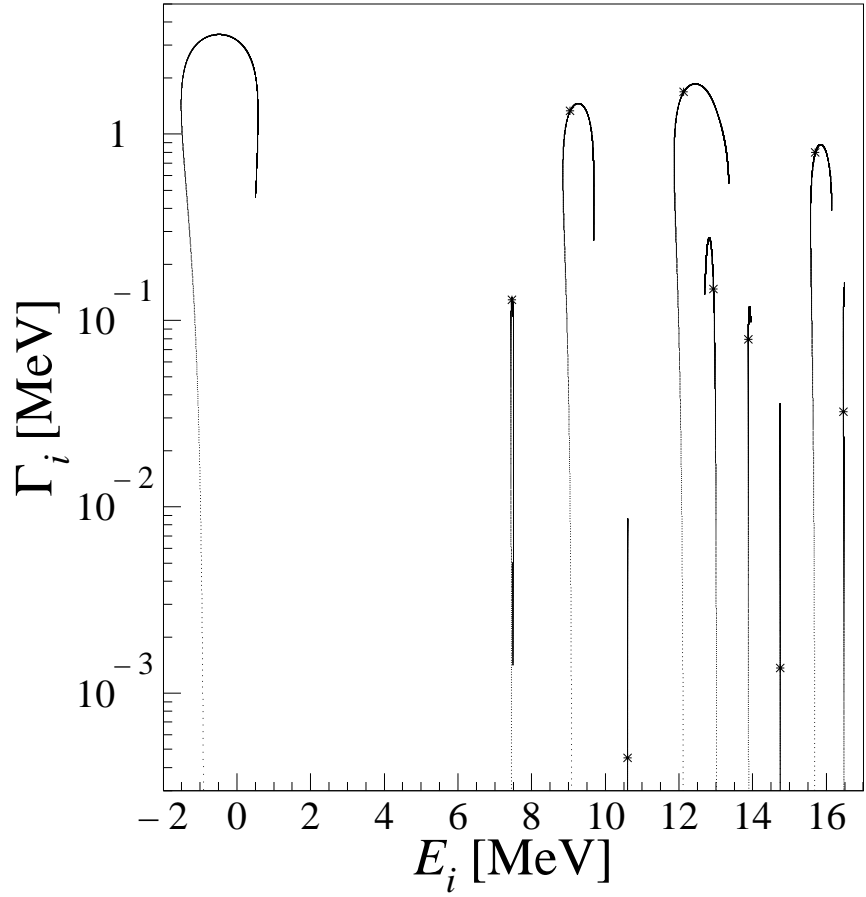}
\end{center}
\vspace*{1cm}
\caption{The eigenvalue picture 
with the energy $E$ of the
particle in the continuum parametrically varied for ten 
lowest $0^+$ states of $^{24}$Mg.
}
\label{paren}
\end{figure}

The resonance states shown in Fig.~\ref{paren1} do not overlap strongly. 
Nevertheless,  the relation between their widths $\tilde \Gamma_k$
and the coupling matrix elements $(\tilde \gamma_k^c)^2$ is far from
being both well defined and energy independent, even in the one-channel case.
In Figs.~\ref{croresen} and  \ref{croresen1}, we show the total
widths $\tilde \Gamma_k $ and   the real and imaginary parts of the 
coupling matrix elements  $(\tilde \gamma_k^c)^2$ for six of these states 
and, additionally,  the partial widths $|\gamma_k^c|^2$ 
(the unambiguous identification of the label for different states can be
obtained from  Fig. \ref{paren}).
All results are for the one-channel case and
therefore $\Gamma_k= |\gamma_k^c|^2$ is assumed in the 
$R$ matrix theory.

The states `1' and `8' are well isolated from
the other ones due to the large distance in energy (state `1') and
small width (state `8'), respectively. For these two states the 
relation $\tilde
\Gamma_k \approx {\rm Re}(\tilde \gamma_k^c)^2$ holds in the whole energy 
region considered.  
The values  $\tilde \Gamma_k $ and  $|\gamma_k^c|^2$ differ from each other,
but show a similar energy dependence (Fig.~\ref{croresen1}).

States `5' and `6' are coming near to one another at an 
energy higher than their position. 
As a consequence, the total widths $\tilde \Gamma_k $
are different from the Re$(\tilde \gamma_k^c)^2 $ that are the real parts of
the coupling matrix elements of the resonance states to the continuum.
They differ also from the 
$|\gamma_k^c|^2$ that are the coupling matrix elements 
of the discrete states to the continuum. The differences are noticeable
in the whole energy region considered, and not only at their 
nearest distance in
energy (Fig.~\ref{croresen}). This happens for the states 
`7' and `10' in a similar manner (Fig.~\ref{croresen1})
as for the states `5' and `6' although the 
distance to the neighbouring states is, in these cases, much larger 
than in the case of two states `5' and `6'.

Figs.~\ref{croresen} and  \ref{croresen1} illustrate that 
the standard one-level 
formula  for the cross section (the Breit-Wigner 
representation)  
can  be applied only for well isolated resonances. 
Only in such a case, unitarity of the $S$ matrix provides a clear relation
between the 
total widths $\tilde \Gamma_k$  and the coupling coefficients between system
and environment. Well isolated resonances appear, however,
seldom in realistic situations. Therefore, a generic relation of the type
$\tilde\Gamma_k = \sum |\gamma_k^c|^2$ (or  
$\tilde \Gamma_k = \sum (\tilde\gamma_k^c)^2$)
does not exist, even in the one-channel case. 
The partial widths $|\gamma_k^c|^2$ of the state $k$ relative to the channels
$c$ loose their physical meaning when the resonance states are not
well isolated. Furthermore, the $\tilde\Gamma_k$ 
and even the $|\gamma_k^c|^2$ are energy dependent.

These numerical results show that, in general, the influence of
the different states onto the properties of the system can not be 
restricted to the small energy region that is determined by 
their energies and widths. This restriction being
one of the basic approximations of $R$ matrix approaches, is  justified 
neither for
bound states lying just below the first particle decay threshold nor for
resonance states at high level density.  

Thus, even though the  SMEC and  the different $R$ matrix approaches start both
from a reliable nuclear structure model, 
the coupling of the resonance states via the continuum of decay channels
is taken into account correctly only in the SMEC.

\subsection{Statistical versus dynamical aspects of resonance states at high
level density}

The statistical properties of $^{24}$Mg are studied in \cite{drokplro}.
As a result,  the dynamics of the $^{24}$Mg nucleus in
the short-time scale is determined by the states at the edges of the spectra
of the parent and daughter nuclei. These states are strongly related to each
other with the result that the corresponding resonance states have short
lifetimes. Randomness in an open quantum system can be found only in the
long-time scale and, even here, only in the one-channel case. 
Since the short-lived and long-lived states 
are created together at avoided level crossings,  both time scales 
exist simultaneously in the  nucleus.  This statement is in agreement with 
experimental results on different nuclei 
of the $sd$-shell, including  $^{24}$Mg.  
The experimental data show  the  interplay of
various reaction times, ranging from the lifetime of the compound nucleus to
the time associated with shape resonances in the ion-ion potentials 
\cite{braun}. For a more detailed discussion see \cite{rep}.

\section{Relation between lifetimes and decay widths} 

The phenomenon of resonance trapping has been discussed  
also in quantum chemistry for unimolecular reactions. 
For illustration, let
us consider here  the  unimolecular decay processes in the
regime of overlapping resonances with the goal to  elucidate how
unimolecular reaction rates depend on resonance widths \cite{miller}.
Using  the definition 
\begin{eqnarray}
k^{\rm eff} = - \frac{d}{dt}\; {\rm ln} \langle \phi(t)|\phi(t)\rangle
\label{eq:ch3}
\end{eqnarray} 
for the decay rate, and
\begin{eqnarray}
\langle \Gamma \rangle = \frac{1}{N} ~\sum_{k=1}^N \Gamma_k ~ \ ,
\label{eq:ch2}
\end{eqnarray} 
where the sum runs over all $N$  resonance states in the
energy region considered, the result  is as follows \cite{miller}:
in all studied cases, the  dependence of the average decay rate on 
$\langle \Gamma \rangle$  for a given energy interval is characterized by 
a saturation curve. In other words:
in the regime of nonoverlapping resonances (the weak coupling regime),
the standard relation between decay rate and $\langle \Gamma \rangle$  holds, 
{\it i.e.}  the unimolecular decay rate is equal to the
resonance width divided by $\hbar$. When, however, the resonance overlap 
increases (the strong coupling regime), the decay rate saturates
as a function of increasing $\langle \Gamma \rangle$.
Identifying the average resonance width $\Gamma^{\rm av}$ with
$\langle \Gamma \rangle$, it was claimed \cite{miller} that the 
fundamental quantum mechanical relation between the
average decay rate and the average resonance width does not hold 
in the strong overlapping limit.

This conclusion is, however, justified only under the assumption of 
uniform level  broadening that makes it possible to  identify
$\langle \Gamma \rangle$  with $\Gamma^{\rm av}$ \cite{romicom}. 
According to the phenomenon of resonance trapping, Eq. (\ref{corr1}), 
the levels are, however, broadened non-uniformly in the overlapping regime
due to the  reordering processes  taking place  
under the influence of the environment into which the system is embedded. 
As a consequence 
\begin{eqnarray}
\sum_{k=1}^M \Gamma_k \gg  \sum_{k=M+1}^N \Gamma_k ~ \ ,
\label{eq:ch10a}
\end{eqnarray}
and $\Gamma^{\rm av}$ is different from $\langle \Gamma \rangle$
in the overlapping region. 

A meaningful
definition of the average width of the long-lived states is \cite{pegoro}
\begin{eqnarray}
\Gamma^{\rm av} =  \frac{1}{N-M} ~\sum_{k=M+1}^N \Gamma_k  ~ \ .
\label{eq:ch10}
\end{eqnarray}
The sum in (\ref{eq:ch10}) runs over the $N-M$ long-lived 
(trapped) states only. These states do not overlap and the
value $\Gamma^{\rm av}$ saturates in the long-time scale.

The saturation of the decay rate in the overlapping regime may be related 
also to the broadening of the widths distribution occurring in this regime 
\cite{miller}. This result is {\it not} in contradiction with the conclusion 
that the saturation is related to resonance trapping. The point is that
resonance trapping creates differences
in the transmission coefficients for the different states that
cause a broadening of the widths distribution \cite{pegoro}. 
Consequently  both, the broadening of the widths distribution and the
saturation of the decay widths in the overlapping regime, can be traced back 
to the same origin, {\it i.e.} to  resonance trapping or, more generally, 
to  avoided level crossings in the complex plane.

It is possible to invert the discussion: it is not the standard relation
between decay rate  and average decay width which ceases to hold
in the overlapping regime. The saturation of the decay rate  is
rather a proof of the formation of different time scales. 
A uniform level broadening does not take place
in the system at high level density, and 
the unimolecular decay rate in the long-time scale is equal to the
average resonance width  $\Gamma^{\rm av}$ divided by $\hbar$.
It should be mentioned here, that also in atomic nuclei a similar saturation
effect is known:  the spreading width  obtained from an analysis of
experimental data on isobaric analogue resonances in different nuclei 
saturates \cite{HarRi}.

We conclude that the standard relation between decay rate and 
average decay width $\Gamma^{\rm av}$  holds
in the regime of overlapping resonances in the long-time scale. 
The point  is that different time scales exist in this regime  that are
caused by resonance trapping, {\it i.e.} by the bifurcation of the widths  
at the   avoided level crossings.

\section{Properties of the broadened   states}

\subsection{Dynamical localization}

An answer to the question
whether  resonance trapping is accompanied by dynamical changes of the 
shape of the system, can not be found directly from studies on 
microwave cavities,
since their shape is fixed from outside.  Nevertheless, a {\it dynamical 
localization} of the wave function density  inside the cavity
may occur. A  study of different wave functions in open microwave cavities 
showed, indeed, that the
localization of  the probability density for short-lived and 
long-lived states inside the cavity
is different. While the short-lived states are localized along particular short
paths related to the position(s) of the attached wave guide(s),
all the long-lived trapped states have pronounced nodal structure 
that is distributed over the whole cavity
\cite{pepirose,napirose}. The long-lived states can be
described well by random matrix theory \cite{shot}.

A similar result has been obtained in calculations for nuclei \cite{radial}. 
The numerical results obtained for the radial profile of partial widths 
of $1^-$ resonance states with 2p-2h structure in $^{16}$O  
show  that, also in this case, resonance
trapping is accompanied by a dynamical localization of the  short-lived
states. In other words: {\it structures in space and time} 
are created that are characterized by a
small radial extension, a short lifetime and a small information entropy 
\cite{radial,ismuro}.

\subsection{Classical description}

As has been discussed in Sect. 3.2, 
resonance trapping is accompanied by the  broadening of some
states when the system is  opened to the continuum. 
An example are the whispering gallery modes that
appear   in microwave cavities  and 
may give an important contribution to the conductance when the cavity is
opened by attaching wave guides to it \cite{napirose}. 

From a physical point of view, most interesting is the following fact.    
Special states of a certain type are characterized by their 
structure that is similar for all the  states of this type. 
For example, all whispering gallery modes are spatially localized
in different groups parallel to each other
and near to the convex boundary of the cavity. 
The states of each group differ by the
number of nodes, but not by the localization region.
They couple therefore  coherently to the continuum of decay channels when the
leads are attached in a suitable manner.  
As a consequence, special states  existing in closed systems  among other 
states,  may become 
dominant by opening the system to a small number of decay channels
\cite{napirose}. 
The widths  of these  states may increase strongly
by trapping other incoherent or less coherent resonance states lying 
in the same energy region. The special states align with the channels while 
the trapped states  decouple more or less completely from the 
environment. Eventually, the properties of the system as a whole   
are determined in the short-time scale mainly by
the special states. Since these states are localized,
they loose partly their wave character, and
it is even possible to describe some properties of the system
by using the methods of classical physics.

The conductance of the cavity is determined by the non-diagonal terms 
of the $S$ matrix. The transmission coefficients 
between channel $n$ and $m$ may be represented by a Fourier transform 
in order to get the length spectra of the quantum mechanical calculations
\cite{napirose}. They are compared to
the histograms of  trajectories  calculated
classically  as a function of the length $L$ of the path
for the same  cavity. In the classical calculations, the trajectories
are obtained from paths of different lengths corresponding to a 
different number of bouncings of the particle at the convex boundary.
The results display a remarkable and surprisingly good
agreement between the quantum mechanical results
of the Fourier analysis and the classical results in spite  of the small value
of the wave vector  of the propagating waves \cite{napirose}.

The correspondence between the quantum mechanical and purely classical results
holds not only in the length scale but also in the time scale: 
varying parametrically the length $L$ of 
the paths causes corresponding  changes in the widths of the special states 
that agree
with the changes of the times for transmission of a classical particle
through the cavity \cite{napirose}. 
Obviously, the correspondence is related to the spatial  localization of 
the whispering gallery modes due to which the wave properties of 
the states are somewhat suppressed. 

These results illustrate the strong correlations which may be induced in the
system due to its coupling to the continuum. The correlations appear in
the long  time scale and, above all,  in the short time scale. 
While the long-lived trapped
states can be described well by random matrix theory, the short-lived
coherent modes are regular \cite{shot}.

\section{Concluding remarks}

All the studies of open quantum systems have shown that the coupling of
the system to the environment may change the properties of the system. 
The changes are small as long as the coupling strength between system 
and environment
is smaller than the distance between the individual states of the 
unperturbed system, i.e. smaller than the distance between the 
eigenstates of the Hamiltonian $H$.
The changes can, however, not be neglected when the coupling to the continuum 
is of the same order of magnitude as the level distance or larger. 
In such a case, the changes can be described neither by perturbation theory
nor by introducing statistical assumptions for the level distribution.  
The point is that non-linear effects become important which cause 
a redistribution of the spectroscopic properties of the system and,
consequently, changes of its properties.  
Under the influence of the coupling to the continuum, level 
repulsion as well as level attraction may appear that are accompanied by
the tendency to form a uniform time scale for the system in the first case,
but different time scales in the second case.

The resonance phenomena are described well by 
two ingredients also at high level density. The first ingredient is
the effective Hamiltonian ${\cal H}$ that contains all the  basic
structure information involved in the Hamiltonian
$H$, i.e. in the Hamiltonian of the corresponding closed system with discrete
eigenstates.  Moreover, ${\cal H}$ contains the coupling matrix
elements  between discrete and continuous states
that account for the changes of the system 
under the influence of its coupling to the continuum. These matrix
elements are responsible
for the non-Hermiticity of ${\cal H}$ and  its complex eigenvalues
which determine not only the positions of the resonance states but
also their (finite) lifetimes.

The second ingredient is the unitarity of the $S$ matrix that has to be
fulfilled in all calculations of resonance phenomena. 
It is taken into account in the unified description of structure and
reaction aspects since any statistical
or perturbative  assumptions are avoided in solving 
the basic equation (\ref{csm1}). 

The studies within the formalism of a unified description of structure and
reaction phenomena show that 
the coupling of the states via the continuum 
induces correlations that are not small. The correlations are important 
especially in the short time scale, but appear also in the long time scale.
The  short-lived states, 
involving  information  on the environment, characterize the system 
far from equilibrium. The long-lived states, however, are described well by
random matrix theory. They are more or less decoupled from the environment.

\newpage

\newpage
\vspace*{3cm}

\newpage
\vspace*{2cm}

\newpage
\vspace*{2cm}

\end{document}